\documentclass[aps,prl,showkeys,superscriptaddress,10pt,nobibnotes,twocolumn]{revtex4-1} %,twocolumn
\usepackage[utf8]{inputenc}
\usepackage[T1]{fontenc}
\usepackage{soul}
\usepackage{dsfont}
\usepackage{amsmath}
\usepackage{graphicx}
\usepackage{bm}
\usepackage[dvipsnames]{xcolor}
\begin{document}

\title{Ferroelectricity and topological vortices from molecular ordering in metal-organic frameworks}

%AUTORI E INFORMAZIONI DI CONTATTO

\author{Francesco Foggetti}
%\email{francesco.foggetti@iit.it} %  se c'è, prima la mail
\affiliation{Quantum Materials Theory, Italian Institute of Technology, Via Morego 30, 16163 Genova, Italy}
\affiliation{Department of Physics, University of Genova, Via Dodecaneso, 33, 16146 Genova GE}

%\author{\ff{...}}
%\affiliation{\ff{...}}

\author{Alessandro Stroppa}
\affiliation{Consiglio Nazionale delle Ricerche, Institute for Superconducting and Innovative Materials and Devices (CNR-SPIN), c/o Department of Physical and Chemical Sciences, University of L’Aquila, Via Vetoio, I-67100, Coppito, L’Aquila, Italy}

\author{Sergey Artyukhin}
%\email{sergey.artyukhin@iit.it}
\affiliation{Quantum Materials Theory, Italian Institute of Technology, Via Morego 30, 16163 Genova, Italy}

%\author{...}
%\email{...}
%\affiliation{...}

%\pacs{PACS numbers:} %Identificano il tipo di articolo, servono sempre?
\newcommand{\dma}{(DMA)Fe$^{II-III}$(COOH)$_3$\;}
\begin{abstract}
% \st{Hybrid inorganic-organic perovskites replace inorganic ions in the usual ABO$_3$ structure by organic molecules, which often results in enhanced structural flexibiliy, rich phase diagrams  and rich functionalities.} 
{Metal-organic frameworks comprehend a wide class of hybrid organic-inorganic materials with general structure A$_m$BX$_n$, with $A$ and $X$ being organic molecules and B a metal cation.  This often results in enhanced structural flexibility and new functionalities. Hybrid perovskites ABX$_3$ are a well-known example.} In an Iron-based perovskites, \dma, dimethylammonium (DMA) molecules are organized in a hexagonal structure \cite{Jain2016}. They are orientationally disordered at high temperatures, but order at around $T=100$~K in a peculiar toroidal pattern. Recent experimental and theoretical study suggest the appearance of ferroelectric polarization in this phase, %\st{but, the microscopic origin has not been elucidated yet}
{although the measured polarization is small, and the mechanism of ferroelectricity is still debated}. We formulate a Landau-type theory that clarifies the connection between the electric polarization, molecular pattern, and distortive modes of the inorganic lattice. 
{We find a remarkable mechanism of improper ferroelectricity, analogue to the trimerization process in inorganic hexagonal ferrites and manganites, but here driven by the ordering of organic molecules in a metal-organic framework.}
Our study reveals an extremely rich phase diagram with the prediction of topological domain walls, where the ferroelectricity arise from tripling the unit cells due to molecular ordering. Wide domain walls with inner structure are predicted.
\end{abstract}

%\keywords{Multiferroics, Magnetoelectric Effect, Magnetic Monopoles, Spin Waves, Non-reciprocal magnons} %Surface acoustic waves, Impedance Microscopy

\maketitle

{\it Introduction} -- %\st{Hybrid organic-inorganic perovskites are a promising realization of perovskite materials in which organic molecules substitute inorganic ions in the usual ABO$_3$ structure, leading to cheaper materials with superior performance or new functionalities. }
{Metal-organic frameworks (MOFs) are hybrid organic-inorganic compounds where the presence of organic molecules enables novel properties, not replicated by any inorganic-based compounds \cite{li2018} and resulting in cheaper materials with superior performance or new functionalities.}
Despite many molecules are polar, observations of ferroelectricity in molecular compounds has been limited until recently \cite{Horiuchi2008,Ghosh2022}. New mechanisms, specific for organic compounds, have been identified, including neutral-to-ionic transition in charge transfer complexes, and hydrogen bond mechanism, as realized in donor-acceptor molecular chain compounds \cite{Li2013}.
Co-DMA metal-organic framework, involving dimethylammonium (DMA) molecules NH$_2$(CH$_3$)$_2$, has recently been discovered to host multiferroic state \cite{Zapf16}. The hybrid nature may combine the advantages of both organic and inorganic semiconductors for optoelectronic applications. For example, a striking and very recent direction is the  possibility is to exploit the chirality of the organic component for obtaining chiral hybrid perovskites through the transfer of the chirality of the molecules to the achiral octahedral layers.\cite{Long2020chiral}
%% can you introduce this reference: https://www.nature.com/articles/s41578-020-0181-5
%%--------SOLVED

A peculiar MOF is \dma , a mixed-iron hybrid inorganic-organic perovskite. {\it Ab-initio} calculations suggested a  polar structure below 100~K with R$3c$ space group \cite{Tang2019}. In this phase, DMA molecules order and the estimated ferroelectric polarization along (111) direction is $\sim$ 7 $nC/cm^{2}$, slightly larger than the experimental measured polarization about 1 $nC/cm^{2}$ \cite{Guo2017}.
% please cite this the experimental paper:
% J. Guo, L. Chen, D. Li, H. Zhao, X. Dong, L. Long, R. Huang, and L. Zheng, Appl. Phys. Lett. 110, 192902 (2017)
%%--------SOLVED
 
However, due to the complexity and size of the system, the exact origin of ferroelectricity has not been clearly identified. It may be related to molecular orientations or charge order. Moreover, ferroelectric order of molecules with magnetic ordering of Fe ions, could lead to a multiferroic state \cite{Zapf16}. Here,  we build a phenomenological modeling based on symmetry considerations  in order to address the microscopic origin  of ferroelectricity in \dma\, and to study possible topological defects.

%to related both magnetic and electric properties \ff{a change of structure from non polar group 167 to the polar one 161 happens at low temperature, this allow for the measured polarization of the order of $nC$}. The presence of Fe in the structure should allow for magnetic properties while ferroelectricity appears to be related to DMA orientation. DMA molecules are polar, but they orientation in an atiferroelectric pattern with neighboring dipoles rotated by 120$^\circ$ in a toroidal-like configuration (Fig. \ref{167}).
%Charge redistribution insi de DMA molecules generates indeed permanent electric dipole moment inside the molecule,

This paper proceeds as follows: the polar R3c structure is considered as a starting point, we then introduce a parent structure and all possible displacive modes that are responsible for the transition to the polar structure. The identified modes act as order parameters for the transition between high and low symmetry structures. Finally a symmetry analysis of the low symmetry structure is performed and a Landau free energy is built from possible low-order invariants.
This allows us to model the interactions between molecular orientation pattern and ferroelectricity. We find a region of model parameters where material behaves as  improper ferroelectric, with the polarization originating from toroidal molecular order. We also predict 3 or 6-fold vortex topological defects, as suggested by the form of the free energy.

\begin{figure}[b]
\centering
\includegraphics[width=\columnwidth]{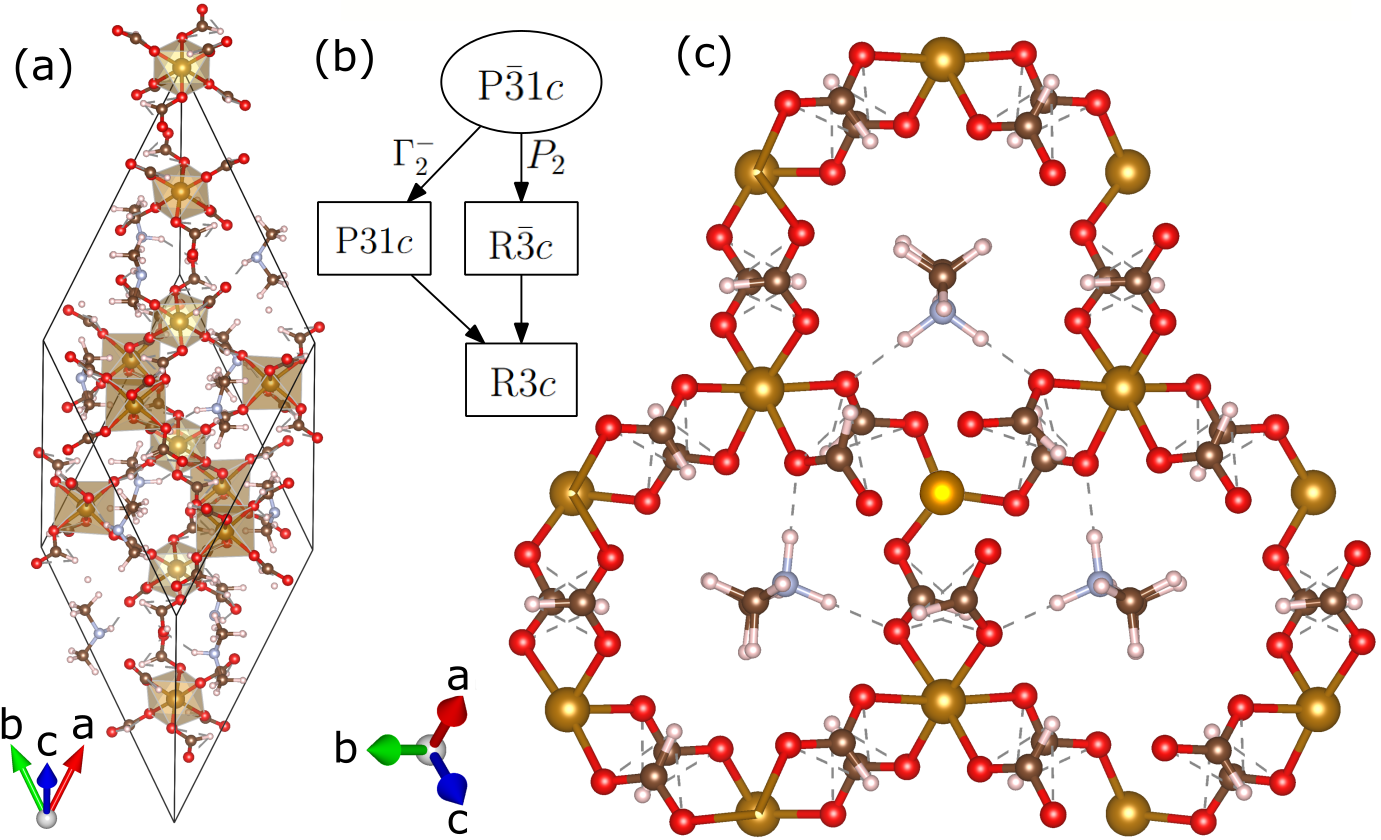}
\caption{\label{fig_163} (a) Rhombohedral unit cell of \dma\, in space group P$\bar{3}_1$c. Fe ions are indicated by light brown balls and oxygens are in red. (b) Symmetry lowering paths, corresponding to different orders in which distortions are removed during the pseudosymmetry search, and the corresponding space groups. The toroidal ordering of molecules ($P_2$ mode) (c) View on the slice of the Fe-MOF along (111) polar direction. DMA molecules reside in circular voids and order in a toroidal fashion. Hydrogen bonds are indicated by dashed lines.}
\end{figure}

{\it Symmetry considerations} -- Our analysis of the system starts with the low-symmetry structure having  the polar space group R$3c$ (\#161 in the International Tables)\cite{Canadillas2012}. The symmetry breaking at low temperatures is due to the ordering on the DMA molecules in a toroidal pattern, and the associated displacements of the neighboring ions, illustrated in Fig.~\ref{fig_163}(c). We thus identify the higher symmetry structure by removing the DMA molecules and looking only at the distortions of the framework. Very recently, Fe-MOF was suggested to have an antiferroelectric state at low $T$ \cite{Canadillas2012}, while indications of ferroelectricity were observed \cite{Guo2017} and theoretical predicted\cite{Tang2019}.
We will see below, that our microscopic model can explain this contradictory results.

We search for pseudosymmetry parent groups using Pseudosymmetry Utility of the Bilbao Crystallographic Server \cite{Capillas2011} and by sequentially removing trimerizing mode P$_2$ and ferroelectric displacements $\Gamma_2^-$, we find a candidate parent structure with space group P$\bar{3}_1$c (\#163 in the International Tables, Fig. \ref{fig_163}(b)). Incidentally, we note that  this structure has also been observed experimentally \cite{Canadillas2012}. By performing mode decomposition \cite{isodistort,Campbell2006} of the low-temperature structure, where the DMA molecules have been removed, with respect to this parent structure, we find the P$_2$ and a polar $\Gamma_2^-$ modes with amplitudes of $2.32$~\AA\, and $0.15$~\AA\, respectively. %\AA does not work in math mode
%GM2- 0.15 AA P2 2.32 AA
P$_2$ mode describes modulated displacements of (COOH)$^-$ organic ligands due to rotating molecular pattern in the original structure, while ferroelectric $\Gamma_2^-$ mode corresponds to polar ionic shifts along the $c$ axis. Fig.~\ref{fig_163}(b) describes how these modes break the symmetry of the parent structure and bring the system into the lower symmetry phase. Table~\ref{tab_inv} shows all possible invariant quantities that we build with $P_2$ and $\Gamma_2^-$ modes, up to the 6$^{th}$ order.
%By looking at the irreducible rion to contextu entering in the Landau Free Energy for the system we consider the generators of group P$\bar{3}_1$c and look at the transformation properties of $\Gamma_2^-$ and P$_2$. By combining the two modes we build the invariant quantities that are symmetric under all the generators of P$\bar{3}_1$c  shown in Table \ref{tab_inv}.

% \begin{table}[t]
% \begin{tabular}{|c|c|c|c|c|}
% \hline
%  &1&$3_{001}^+$&$2_{1\bar 1 0}$&$\bar 1$\\\hline
%       $P_z$&$+$&$+$&$-$&$-$\\\hline
% $\phi$&$\phi$&$\phi+\frac{2}{3}\pi$&$-\phi+\frac{3}{2}\pi$&$\phi+\pi$\\\hline
% %       $\begin{pmatrix}Q_x\\Q_y \end{pmatrix}$&$\begin{pmatrix}1&0\\0&1 \end{pmatrix}$&$\begin{pmatrix}-\frac{1}{2}&-\frac{\sqrt{3}}{2}\\\frac{\sqrt{3}}{2}&-\frac{1}{2} \end{pmatrix}$&$\begin{pmatrix}0&-1\\-1&0 \end{pmatrix}$&$\begin{pmatrix}-1&0\\0&-1 \end{pmatrix}$\\\hline
% \end{tabular}
% \caption{\label{rep} Transformation properties of $P_z$ and $\phi$ under the symmetries of group P$\bar 3_1$c (\# 163) .}
% \end{table}

\begin{table}[t]
\begin{tabular}{|c|c|}
\hline
Degree&Invariants\\\hline
       2&$Q_x^2+Q_y^2,\;\;P_z^2$\\\hline
       3&$3Q_x^2 Q_y-Q_y^3$\\\hline
       4&$P_z^4,\;\;Q_x^4+2Q_x^2Q_y^2+Q_y^4$ \\
        &$P_z^2(Q_x^2+Q_y^2),\;P_z(Q_x^3-3Q_x Q_y^2)$\\\hline
        \raisebox{5pt}{\vphantom{M}}
       6&$(3Q_x^2Q_y-Q_y^3)^2,\;\;P_z^6,\;\;P_z^4 (Q_x^2+Q_y^2),$\\ &$P_z^2(Q_x^2+Q_y^2)^2,\;\;(Q_x^2+Q_y^2)^3,$\\
&$P_z^3(Q_x^3-3Q_xQ_y^2),\;\;P_z(Q_x^2+Q_y^2)(Q_x^3-3Q_xQ_y^2)$\\
        \hline
\end{tabular}
\caption{\label{tab_inv} Invariants for symmetry group P$\bar 3_1$c (\# 163). $Q_x$ and $Q_y$ are the components of the P$_2$ mode while $P_z$ refers to $\Gamma_2^-$ mode.}
\end{table}

\begin{figure}[t]
    \centering
    \includegraphics[width=
\linewidth]{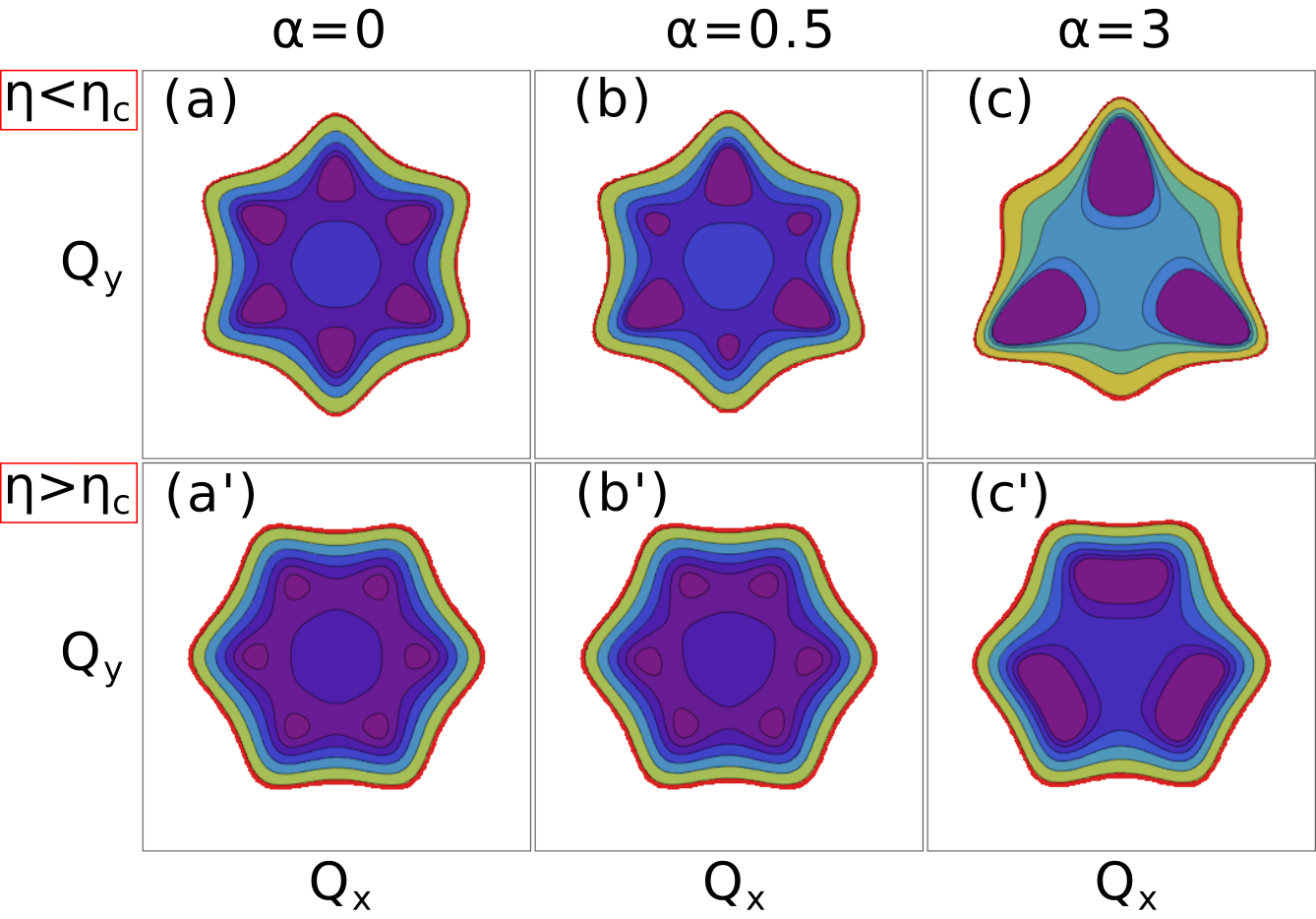}
    \caption{\label{fig_freeReduced} Different energy densities for different values of $\alpha$ and $\eta$. (a-c) $\eta<\eta_c$ case. (a) In $\alpha=0$ there are six equivalent minima, (b) three minima become less deep until they become maxima (c) for the free energy. (a'-c') $\eta>\eta_c$ case. (a') in $\alpha=0$ six minima are present, the angular position of the minima is rotated with respect to the (a) case due to the $\eta\sin^2 3\phi$ term dominating over the $\cos^2 3\phi$. (b') the position of the minima changes, deforming the structure, pair of minima start to merge. (c') The six paired minima merged in three shallow minima.
    }
\end{figure}

{\it Model -- } We use the mode amplitudes for $\Gamma_2^-$ and P$_2$ modes as order parameters to construct a Landau-type free energy of the system,
\begin{equation}
\begin{split}
F=\gamma P_z^2+&(-AQ^2-BQ^4+CQ^6)+\alpha Q^3\sin 3\phi\\
&+\eta Q^6\sin^2{3\phi}+ V(P_z,Q,\phi),
%+\beta P_Z 3Q^3\cos 3\phi
\end{split}
\end{equation}
with ${\bf Q}=(Q_x,Q_y)=(Q\cos\phi,Q\sin\phi)$ parametrizing the P$_2$ mode and $P_z$ standing for the ferroelectric polarization induced by $\Gamma_2^-$ mode. %Table \ref{rep} describes the transformation properties of $P_z$ and $\phi$ under the symmetry operations of the system.
$P_z$ is bounded to be small close to the phase transition, therefore we neglect high order terms in $P_z$ from Table~\ref{tab_inv}.
The term $V(P_z,Q,\phi)$ describes the coupling between the distortive mode and the ferroelectric polarization.
The term with $\gamma>0$ describes a stable polar mode with ferroelectric polarization $P_z$, while the term in parentheses represents a Mexican hat potential. The terms with $\alpha$ and $\eta$ produce a triangular warping of the rim of the Mexican hat, generating three degenerate minima similar to those in Fig.~\ref{fig_freeReduced}(c,c'). These terms take into account the 3-fold symmetric anisotropy and are related to the symmetry of the framework that contains the molecules.
Microscopically, these terms correspond to the anisotropy energy of the molecule, i.e. to the energy cost of 60$^\circ$ rotation away from the easy direction. The terms with $\alpha$ and $\eta$ result from interactions between the molecule and neighboring ions, cf. Fig.~\ref{fig_163}(c). The interaction between the polarization $P_z$ and the $Q$ mode is given by the term
\begin{equation}
    V(P_z,Q,\phi)=\beta P_z (Q_x^3-3Q_x Q_y^2)=\beta P_z Q^3\cos 3\phi,
\end{equation}
%\ff{high order here too? Add it or not? It will be neglected anyway}
that is another invariant shown in Table \ref{tab_inv}. In principle a term $P_zQ^5\cos 3\phi$ should also be considered, but $Q$ is small close to the phase transition and the cosine term is the same as the 3$^{rd}$ order term, therefore this term does not change the angular structure of the minima and, therefore, it  can be neglected. We observe that polarization changes sign under 60$^\circ$ rotation of the vector $\bf Q$, i.e. a rotation of all the molecules by $60^\circ$.
%We see from this term that, for it to be invariant, polarization must change sign under $\pi/3$ rotations of $\vec Q$ (Fig. \ref{rot}) \ff{this figure takes just space, you can see the change in sign of P with the other fig. as well}.

% Of particular importance is the invariant that couples the polarization and displacive mode, in the considered case polarization is along $c$ axis, we call this invariant I
% \begin{equation}
% I=P_z (Q_x^3-3Q_x Q_y^2)=P_z Q^3\cos 3\phi
% \end{equation}
% where $P_z$ is the electric polarization along $c$ axis while $Q_x$ and $Q_y$ are the $x$ and $y$ component of the displacive P$_2$ mode,

%\begin{equation}
%I=P_z Q^3 (\cos^3\phi-3\cos\phi\sin^2\phi)=P_z Q^3\cos 3\phi
%\end{equation}
% \begin{figure}[t]
%     \centering
%     \includegraphics[width=5cm]{./fig/Prot.png}
%     \caption{
%     \label{fig_rot} Polarization along $c$ axis changes sign upon a $\pi/3$ rotation}
% \end{figure}
We now minimize the free energy $F$ with respect to $P_z$, which gives
\begin{equation}
    \begin{split}
    P_z|_{min}=-\frac{\beta Q^3}{2 \gamma}\cos 3\phi,
    \label{eq_Pz}
    \end{split}
\end{equation}
and at this optimal $P_z$ the free energy is
\begin{equation}
\begin{split}
    F=-\frac{\beta^2Q^6}{4\gamma}\cos^23\phi+&\alpha Q^3\sin 3\phi+\eta Q^6\sin^23\phi\\-AQ^2-&BQ^4+CQ^6.
    \label{eq_F6}
\end{split}
\end{equation}
%where we did not rewrite the Mexican hat term that will not be considered in the upcoming analysis.
The condition $C>\beta^2/4\gamma$ must be satisfied to keep the free energy positive at large $Q$.
%%%%%%%%%%%%%%%%%%%%%%%%%%%%%%%%%%%%%%%%%%%%%%%%%%%%%%%%%%%%%%%%%%%%%%%%%%%%%%%%%%%%%%%%%%%%
%%%%%%%%%%%%%%%%%%%%%%%%%%%%%%%%%%%%%%%%%%%%%%%%%%%%%%%%%%%%%%%%%%%%%%%%%%%%%%%%%%%%%%%%%%%%

\begin{figure*}[ht]
    \centering
    \includegraphics[width=.9\linewidth]{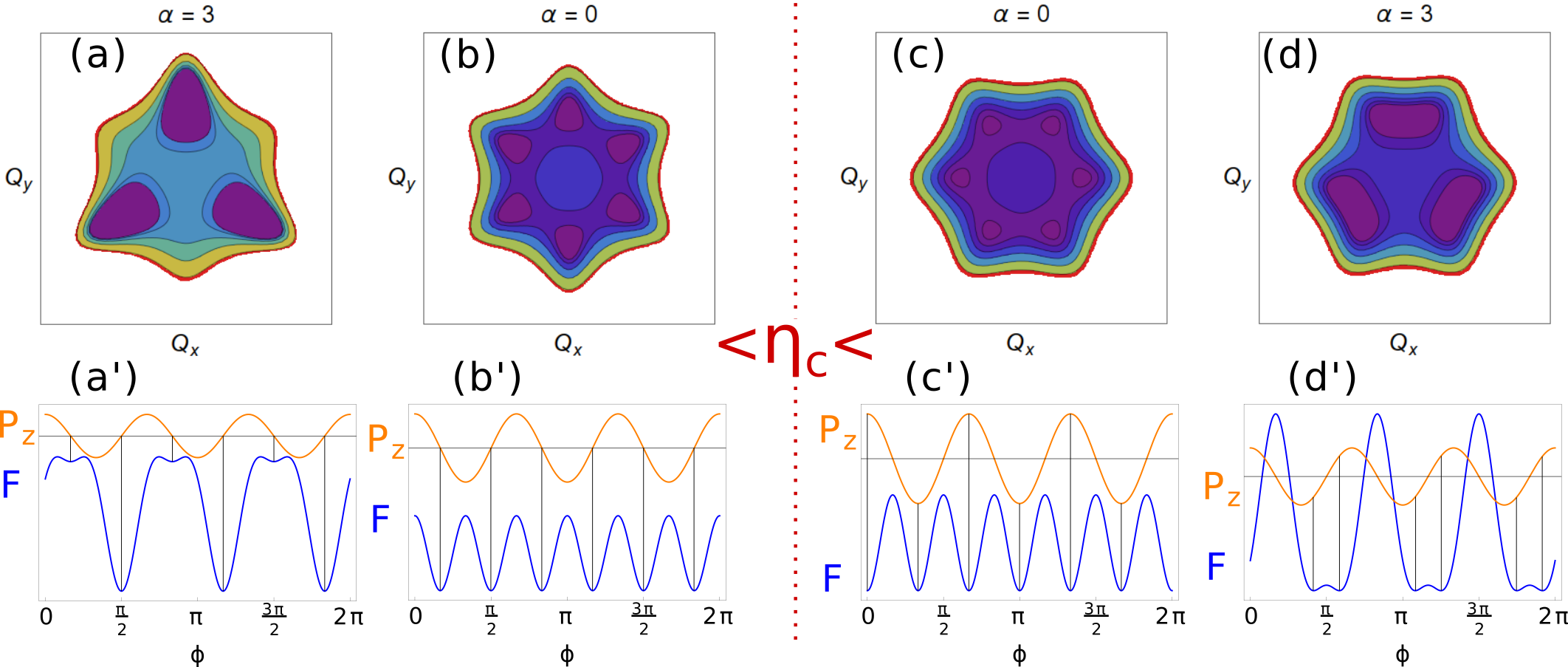}
    \caption{(a-d) Free energy density for different values of $\alpha$ and $\eta$. (a'-b') polarization profile and free energy as a function of the angle $\phi$ for different values of $\alpha$ and $\eta$. In $\eta<\eta_c$ (left) the polarization is zero in correspondence of the minima of the free energy. Solid lines in (a',b') connect the minima of $F$ to the corresponding value of $P_z$. In $\eta>\eta_c$ the polarization is non zero at the minima of the free energy, it has the maximum value (in modulus) at $\alpha=0$ and decreases with growing $|\alpha|$.}
    \label{fig_AlphaEta}
\end{figure*}

We notice from Eq.~(\ref{eq_F6}) that $\eta$ and $\beta$ term are of the same order in $Q$ and can compete with with each other. A critical value $\eta_c=-\beta^2/4\gamma$ defines two cases in which either the sine or cosine term dominates in the free energy. Fig.~\ref{fig_freeReduced} shows how the structure of the minima changes in the two cases. Supplementary Figures~\ref{fig_supplementary_etaSmall} and \ref{fig_supplementary_etaBig} show in greater detail how the free energy density changes with the parameters $\alpha$ and $\eta$.

To simplify the analysis we redefine $F$ by dividing by the coefficient of the cosine term
\begin{equation}
\begin{split}
    F=(\eta+1) Q^6\sin^23\phi+&\alpha Q^3\sin 3\phi-\\-AQ^2-&BQ^4+CQ^6,
    \label{eq_FFinal}
\end{split}
\end{equation}
where all the coefficients in the equation have been adequately rescaled and $\eta_c=-1$. We show in Fig.~\ref{fig_AlphaEta} the free energy density from Eq.~(\ref{eq_FFinal}) and the polarization $P_z$ from Eq.~(\ref{eq_Pz}) for different values of $\alpha$ and in the two cases $\eta>\eta_c$ and $\eta<\eta_c$.

We now discuss the role of $\alpha$ and $\eta$ terms. The term $\alpha Q^3\sin3\phi$ has three minima while the terms $Q^6\cos^23\phi$ and $\eta Q^6\sin^23\phi$ have six minima. As we vary $\alpha$, the total number of minima in the free energy changes. Small values of $\alpha$ will favor the 6$^{th}$-order terms and the free energy will have six minima (Fig.~\ref{fig_freeReduced} (a,b) and (a'b')). As $|\alpha|$ grows the $\alpha Q^3\sin3\phi$ term dominates over the 6$^{th}$-order terms forcing the free energy into a three-minima configuration. A change of sign in $\alpha$ produces a 60$^\circ$ rotation of the minima of the free energy (cf. supplementary figures~\ref{fig_supplementary_etaSmall} and \ref{fig_supplementary_etaBig}). The coefficient $\eta$ controls the competition between the $-Q^6\cos^23\phi$ term and the $\eta Q^6\sin^23\phi$ term and selects the position of the free energy minima in the low-$\alpha$ range (cf. Fig.~\ref{fig_freeReduced}(a,a')).
%\ff{We did not contract the two terms into one to explicitly relate the position of the minima with either the cosine or sine term}.
The $\eta$ coefficient select the cases where a finite polarization can be present in the material. Figure~\ref{fig_AlphaEta} (a'-d') shows the polarization $P_z$ (in orange) computed with Eq.~(\ref{eq_Pz}) as a function of the angle $\phi$ and the corresponding value of the free energy (blue) for a fixed $Q$ and the same angle. We see that, if $\eta >\eta_c$ and for small values of $\alpha$, the polarization has a finite value at the minima of the free energy. On the other side, if $\eta<\eta_c$ the polarization is always zero at the free energy minima, that correspond to the structural domains. Although the polarization $P_z$ is zero at the minima, it has non-zero values in the domain walls, across which ${\bf Q}$ interpolates between the neighboring minima. To summarize, in the case $\eta>\eta_c$, the system exhibits ferroelectric domains, while for $\eta<\eta_c$ it has ferroelectric domain walls.
 %We see that for $|\alpha'|>1$ the energy surface exhibits a three-fold structure with three domains separated by domain walls, in this case the polarization has zero value at $\vec Q$ corresponding to the minima of the free energy. As $\alpha'$ changes sign we observe the splitting of the three-fold structure into six-fold with the appearance of a total of six minima at $\alpha'=0$. In this scenario the polarization assumes non-zero values in correspondence with these new minima to the point of reaching its maximum value when $\alpha'=0$ and the free energy shows six equivalent minima. As $\alpha'$ moves away from zero we return to the initial situation with three minima where $F(\vec Q)$ is rotated by $\pi$ with respect to $F(\vec Q)$ for $\alpha<0$.

\begin{figure}[hb]
    \centering
    \includegraphics[width=.8\linewidth]{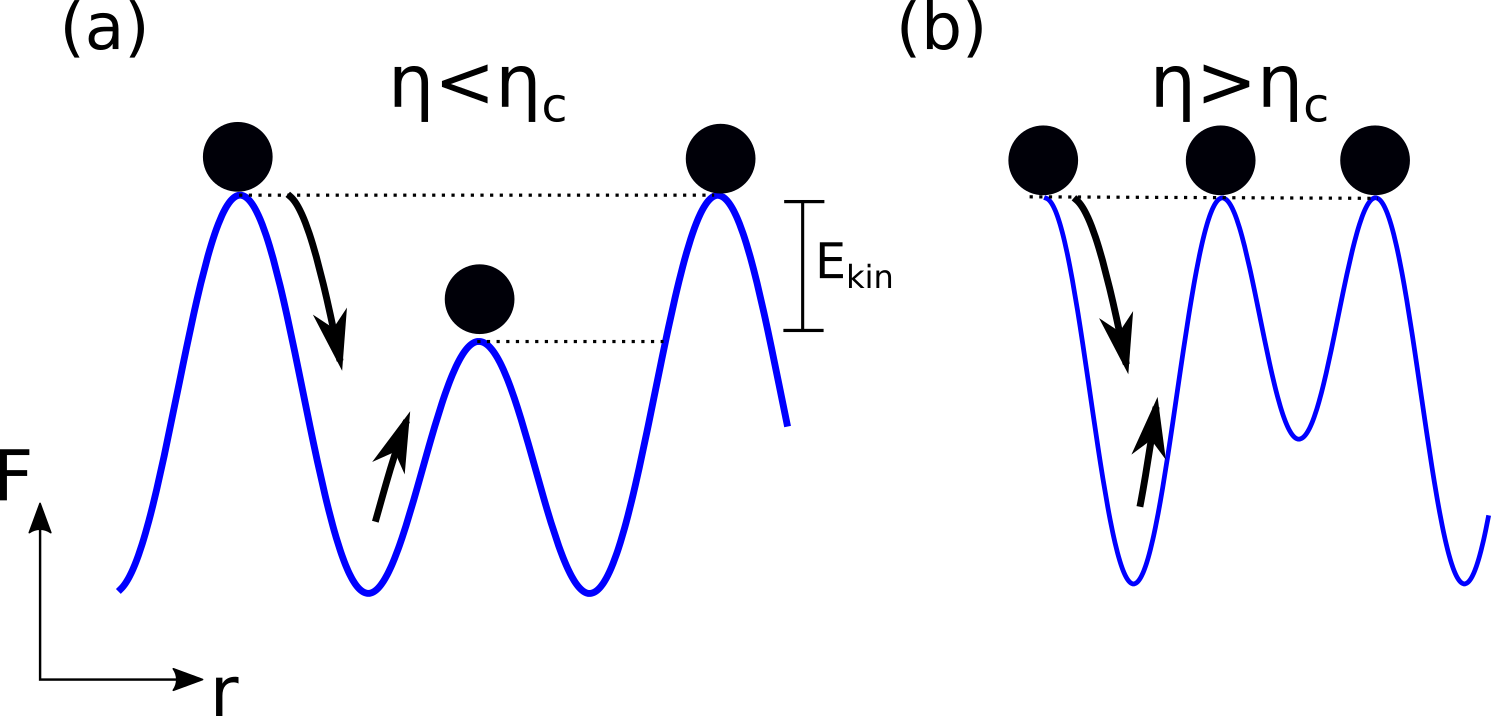}
    \caption{Classical particle in a flipped free energy potential (blue curve) in the case $\eta<\eta_c$(a) and $\eta>\eta_c$ (b). (a) on top of the lower hill the particle has lower kinetic energy than in the minimum and it spends a greater amount of time there. Equivalently, $Q$ stays in a metastable state, which results in the inner structure of the domain wall. (b) all hills have the same height and no metastable state is present.}
    \label{fig_flipped}
\end{figure}

{\it Inner structure of domain walls -- }
%eta gives two cases, only eta > crit da struttura interna. Fix notation Q(x), x not defined. Not clear trajectories, minima, vortexes etc... Refer to figure phase diag e metti figure strutt. interna
The minima of the free energy represent the possible orientation of DMA molecules. At small $\alpha$ all the barriers along the rim of the Mexican hat have the same height, and the six minima are energetically equivalent. 

\begin{figure}[ht]
    \centering
    \includegraphics[width=\linewidth]{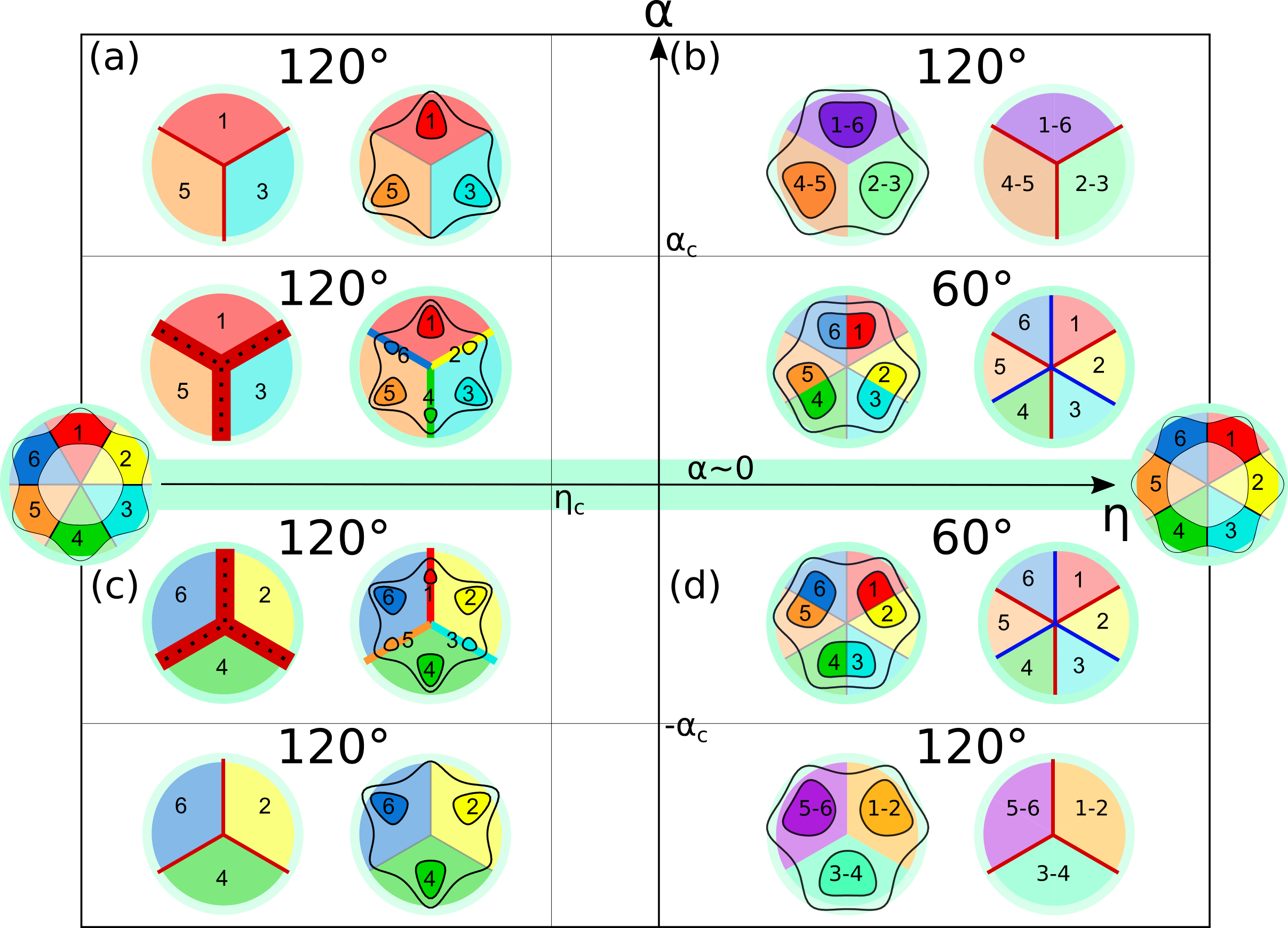}
    \caption{Summarized phase diagram in the parameter space [$\eta$,$\alpha$] computed with the free energy from Eq.~(\ref{eq_FFinal}). At $\alpha=0$ six equivalent minima give rise to 6-fold vortices, shown at the endpoints of the horizontal $\eta$ axis. The phase diagram shows a schematic profile of the energy density (with contours in black) for every phase in the parameter space [$\eta$,$\alpha$] and the corresponding structure of domain walls and vortices. The value $\eta_c$ separates two regimes where six-minima turn into three with different mechanisms. The value $|\alpha_c|$ selects the threshold on the $\alpha$ axis, after which there are only three minima in the free energy and no internal structure is present in the domain walls.}
    \label{fig_phDiag}
\end{figure}

Six domains with different $\bf Q$ orientations are possible and, when the domains meet at the same line (vortex line), an hexagonal vortex defect is formed \cite{Artyukhin2014}. A cut perpendicular to the vortex line is shown in Fig.~\ref{fig_phDiag}~(d, top right). 
%vortex= 6 domains, every domain has one orientation of DMA, every domain is connected to all the others at the center of the vortex, and from one domain to the nearby one there is a step in the rotation of the molecule by 60° Vortez line=stacking of vortex centers in a tube like structre (like smyrmions)
Figure~\ref{fig_phDiag} shows a schematic phase diagram of our model in the [$\eta,\alpha$] space. When moving along the $\alpha$ axis, at $|\alpha|>\alpha_c$, we obtain a phase with three minima, corresponding to three possible domains separated by domain walls.

The inner stucture of domain walls here is connected to three of the six minima going up in energy, thus making the corresponding states metastable.
Once we discuss the domain walls, \emph{i.e.} configurations with spatially varying order parameter, we must complement our free energy density from Eq.~(\ref{eq_FFinal})
\begin{equation}
F_{\mathrm{tot}}=\int dr\, \left(\frac{m}{2}[(\nabla Q_x)^2+(\nabla Q_y)^2]+F(Q)\right),
\label{eq_F_tot}
\end{equation}
with the first term being the stiffness term that penalizes order parameter variations. Then the problem of describing the structure of the domain walls, \emph{i.e.} minimizing Eq.~\ref{eq_F_tot}
is equivalent to the problem of determining the trajectory $Q(r)$ (with $\phi=\phi(r)$) of a classical particle in a flipped potential $\mathcal{V}=-F$, \emph{i.e.} minimizing the Lagrangian $m/2\, \dot{x}^2-\mathcal V$. In the case $\eta<\eta_c$ the minima that are less shallow in the free energy (cf. Fig.~\ref{fig_freeReduced}(b) ) are maxima of lower height in the classical analogue (Fig.~\ref{fig_flipped}(a)). The particle starts at the top of the potential (minimum of the free energy) with zero velocity and it passes the neighboring peak (local minimum higher in energy) with a non zero velocity, spending a finite amount of time near the peak. This second peak represents an ``unfavored domain'' inside a domain wall. The time that the particles spends at the smaller peak is proportional to the finite width of this inner domain.

\begin{figure}[htb]
    \centering
    \includegraphics[width=\linewidth]{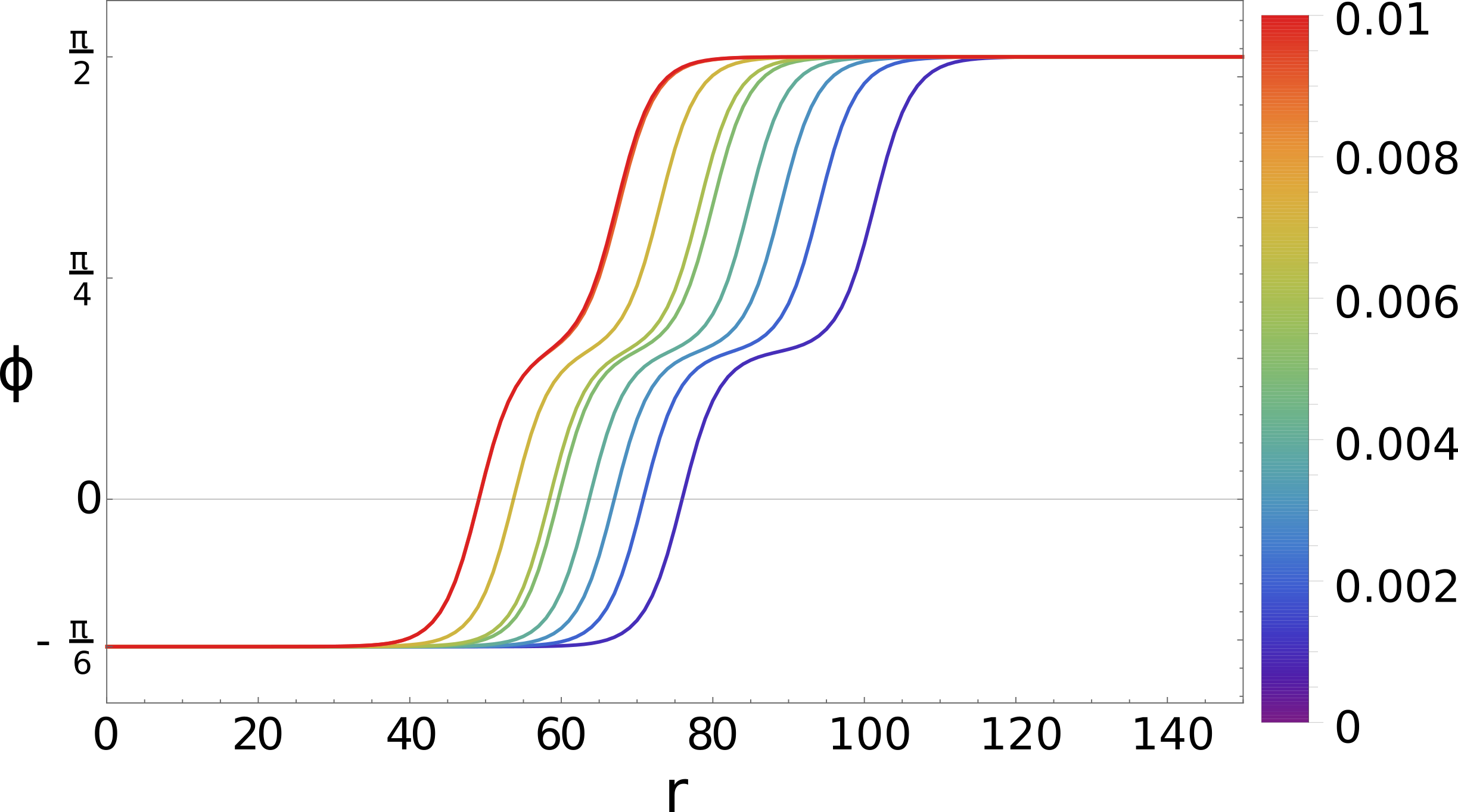}
    \caption{Spatial profile of nested domain walls. The trimerization angle profile $\phi(r)$ represents the structure of the wall, and the inner structure, corresponding to a metastable phase, is present between the two domains, identified by $\phi$ values of $\pi/2$ and -$\pi/6$. The domain wall profile is computed for different values of $\alpha$, which are encoded by the color. The length scale on the horizontal axis (arb. units) is set by the order parameter stiffness, represented by the parameter $m$ in the free energy.}
    \label{fig_metas}
\end{figure}

Domain walls with inner structure, or {\it nested domain walls}, are not commonly observed, therefore the properties of these objects have not yet been characterized. This in turns, clearly suggests a rich phase diagram of these complex hybrid materials, including the possibility of electric skyrmions,\cite{Bostrom2021hybrid} therefore calling for further experimental and theoretical investigations. The phase diagram in Fig.~\ref{fig_phDiag} helps identify the phases where such nested domains walls are expected. Fig.~\ref{fig_metas} shows, for different values of $\alpha$, the presence of inner structure in the domain walls.

In the case $\eta>\eta_c$ all minima of the free energy are equally deep thus, in the classical analogue, all the peaks have the same height and no internal structure is expected (cf. Fig.~\ref{fig_flipped}(b)).
In the supplementary Figure~\ref{fig_supplementary_traj} we report the trajectories $Q$ computed for both $\eta>\eta_c$ and $\eta<\eta_c$. In summary, 
as discussed above, several regimes are possible within the presented theory, such as ferroelectric domains and non-ferroelectric domains with ferroelectric domain walls.
%The detailed realization of one type of domain or the other may depend on the experimental conditions. 
Accurate density functional theory calculations for different compounds will allow to estimate the model parameters and position the materials on the phase diagram (cf. Fig.~\ref{fig_phDiag}).
% Disentangling the different regimes and to estimate the parameters of the model are extremely complex for the present compound, and such as study goes beyond the purpose of the present paper. 
 %The term with $\alpha'$ can be regarded as an electric field, that disfavors the domains of one of the polarizations. Therefore, at $\alpha'>0$ three of six domains become unstable and shrink, which corresponds to the trajectory $Q(x)$ not visiting the local minimum of the potential, and deforming more and more towards the center $Q=0$ as $\alpha'$ is increased. Since there are no local minima corresponding to the same $\phi$ and different $r$, the order parameter trajectories $Q(x), \phi(x)$ can be parametrized as $\phi(x), Q(\phi(x))$. The minimization problem is then cast into the one for $\phi(x)$ and the corresponding $Q(\phi)$ trajectories can be thought of as trajectories of a particle on an inverted potential Fig.~2(top) \ff{figure?}. The particle then starts at the top of the mountain (minimum of the actual potential) with zero velocity, and as it passes the neighboring peak, lower than the one it has started from for $\alpha>0$, the velocity at the peak is non-zero, which gives a finite width to the part of the domain wall that essentially represents the "unfavored domain". This inner structure of the domain walls may be observed experimentally and there may prove useful for applications. \ff{phase diagram?}

%{\it DFT simulations of the energy landscape --} As discussed above, several regimes are possible within the presented theory, such as ferroelectric domains and non-ferroelectric domains with ferroelectric domain walls. In order to identify the phase \dma\, realizes, we estimate the parameters of the theory using first-principles simulations. Nudged elastic band calculations were performed using Quantum Espresso software, with the initial point in the low-$T$ experimental structure \cite{Canadillas2012} and the images along the path obtained by rotating the DMA molecules in 10 steps, by the total angle of $2\pi/3$. The unit cell contained 222 ions and 2 layers of Fe-centered octahedra. The energy cutoff of 90~Ry and SSSP pseudopotentials were used. Only the $\Gamma$ point was used to sample the kinetic energy. LDA+U with Hubbard $U=4$~eV was used for Fe $d$ electrons. The middle NEB image was used as a climbing image. The overall barrier of 1.5~eV per 222 ion cell was found at $\phi=\pi/3$. \ff{asymmetric barrier shows we're in the case eta bigger than eta crit?}

{\it Conclusions --} In this study we built a Landau theory of phase transition describing the molecular ordering and ferroelectric polarization {in MOFs, in particular in the}   perovskite-based hybrid inorganic-organic  %\st{and, in particular,} 
\dma . The results suggest that ferroeletricity can arise from tripling of the unit cell due to molecular ordering. Wide composite domain walls, with the metastable states realized inside the domain walls, are predicted.  We find vortex line defects, where three of six domain walls merge. Our study highlights the complex and rich phase diagram due to the interplay of molecular and framework degrees of freedoms 
thanks to the dual nature of {MOFs}. %\st{hybrid perovskites}
%First-principles calculations for Fe-MOF \footnote{to be published separately} suggest a small polarization, also found experimentally \cite{Guo2017}. 
We hope the our work will motivate experimental and {\it ab-initio} investigations of this mechanism which may be more generally found in {MOFs}. %{hybrid inorganic-organic perovskites}.
%First-principles estimates suggest high barrier, narrow DWs and small polarization in Fe-MOF.

%The presented model describes improper ferroelectricity in \dma, driven by the order of the organic molecules. We also discussed topological defects (hexagonal and triangular vortexes) possible in this model

%We performed a symmetry analysis and derived a Landau-type theory describing molecular ordering and ferroelectric polarization in perovskite-based MOFs. The results suggest a new mechanism of improper ferroelectricity in these materials, that is related to unit cell tripling molecular order. The symmetry aspects are reminiscent of those in hexagonal manganites and ferrites, but the theory has a additional term, that makes domains of opposite polarizations inequivalent. When the term is small, wide composite DWs, hosting the intermediate metastable phase, appear. First-principles estimates suggest high barrier, narrow DWs and small polarization in Fe-MOF.

% \sa{As seen in Fig.~3(b) of \cite{Ma2019}, a pronounced pyroelectric peak appears at the phase transition temperature $T_C\sim 160$~K for the Co-MOF, which yields a relatively large electric polarization ($P\sim 0.1\mu$C/cm$^2$). In contrast, as shown in Fig.~2(d) of \cite{Ma2019}, there is no clear pyroelectric peak at the phase transition for the Fe-MOF and thus the electric polarization is nearly zero.}
\bibliography{organicFE}

\widetext
\newpage
\renewcommand{\thefigure}{S\arabic{figure}}
\setcounter{figure}{0}
\begin{center}
\textbf{\large SUPPLEMENTARY INFORMATION}
\end{center}

\begin{figure}[ht]
    \centering
    \includegraphics[width=\linewidth]{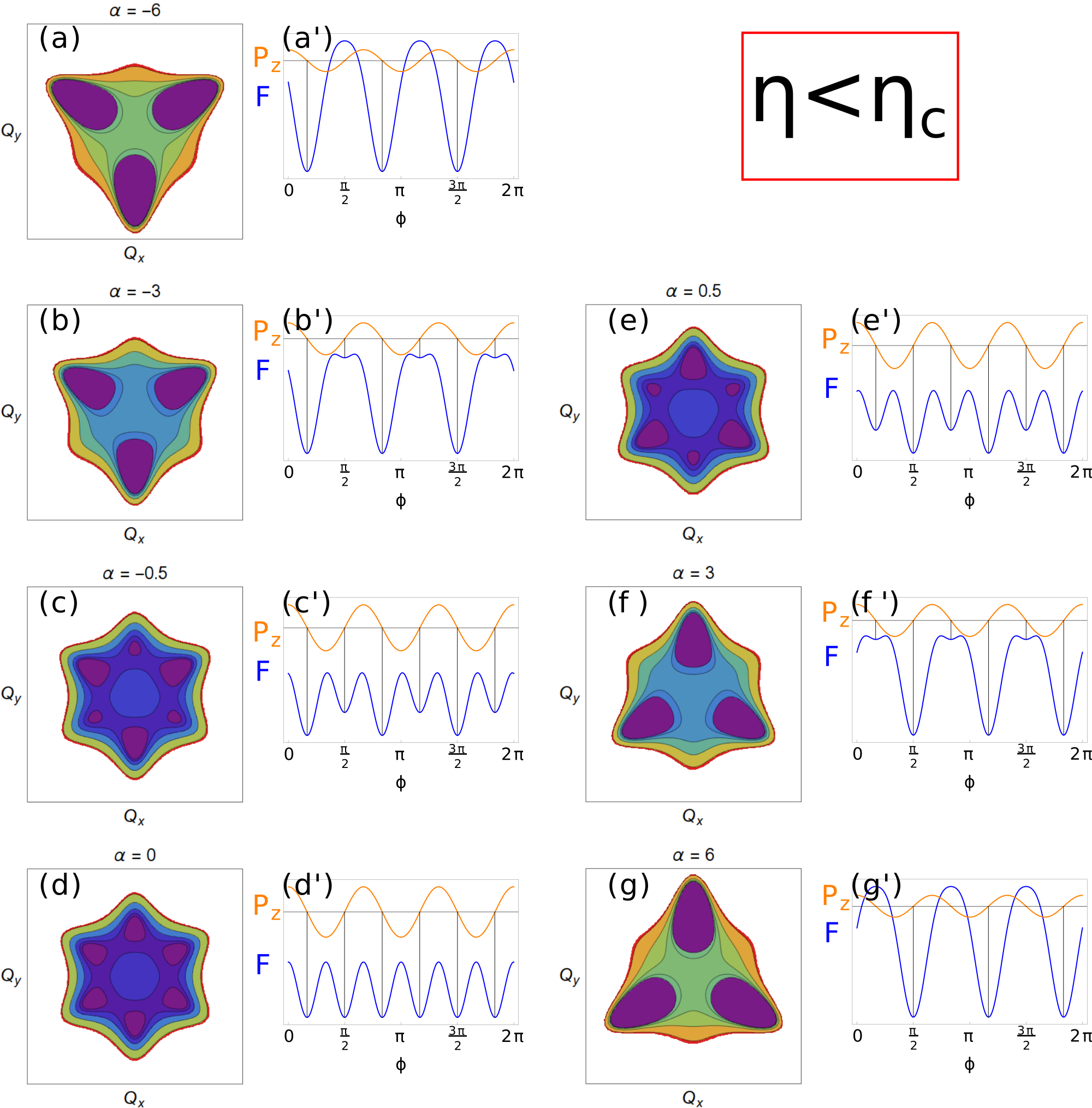}
    \caption{(a-e) Free energy surface in the case $\eta<\eta_c$ for different values of $\alpha$. (a'-e') Free energy $F$ and polarization $P_z$ as a function of the angle $\phi$. Solid lines are a guide to the eye. In correspondence of the minima of $F$ the polarization is zero for all values of $\alpha$. When $\alpha$ changes sign the position of the minima is rotated by 60$^\circ$.}
    \label{fig_supplementary_etaSmall}
\end{figure}

\begin{figure}[ht]
    \centering
    \includegraphics[width=\linewidth]{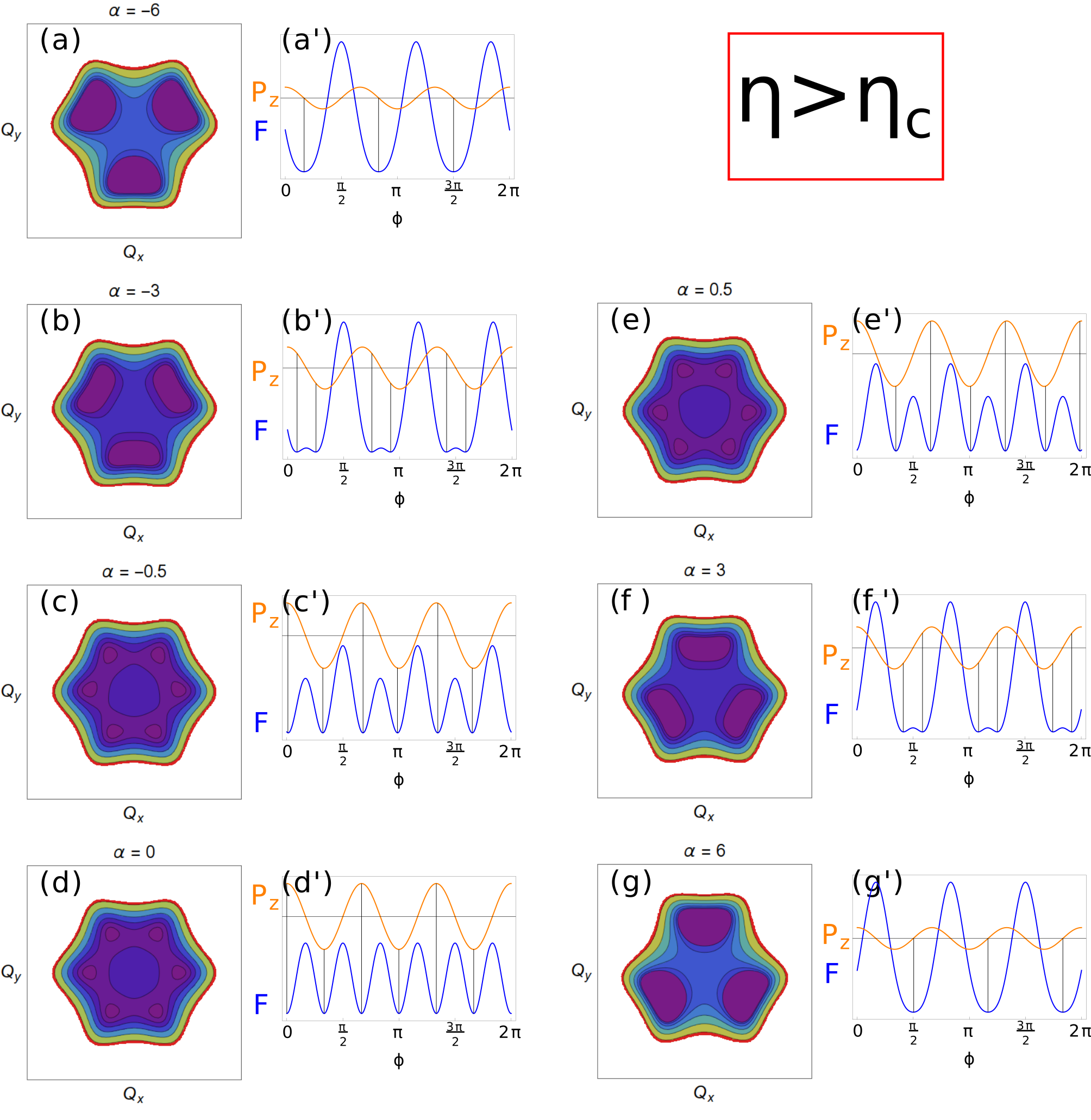}
    \caption{(a-e) Free energy surface in the case $\eta>\eta_c$ for different values of $\alpha$. (a'-e') Free energy $F$ and polarization $P_z$ as a function of the angle $\phi$. Solid lines are a guide to the eye. In correspondence of the minima of $F$ the polarization is non-zero for small values of $\alpha$. When $\alpha$ changes sign the position of the minima is rotated by 60$^\circ$. }
    \label{fig_supplementary_etaBig}
\end{figure}

\begin{figure}[ht]
    \centering
    \includegraphics[width=.8\linewidth]{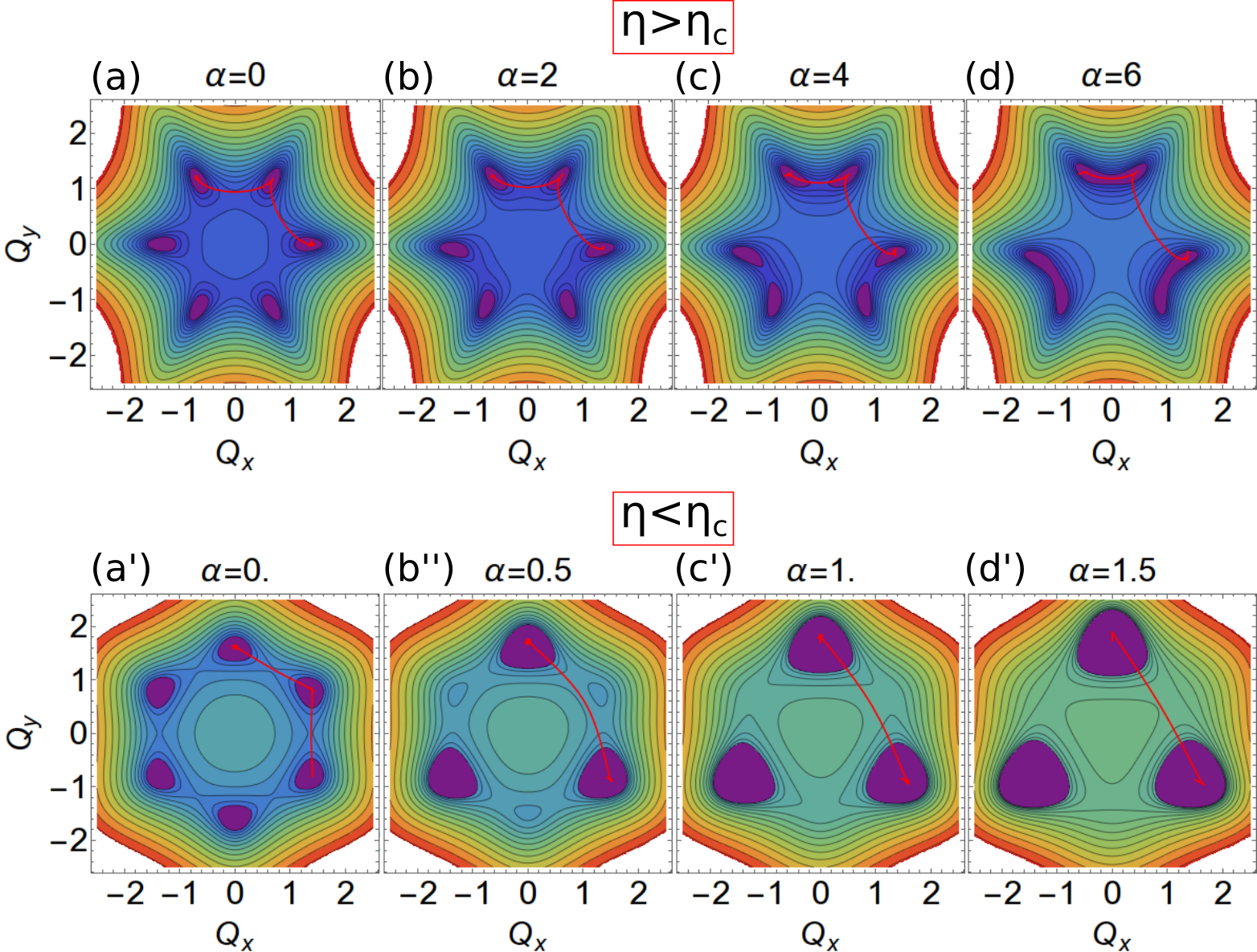}
    \caption{Computed trajectories in the classical problem with flipped potential $\mathcal{V}=-F$, from a minimum to the equivalent one defined by a rotation of 120$^\circ$ degrees. (a-d) $\eta>\eta_c$ case, the transition from 6 to three minima happen by merging of two adjacent minima. The trajectory of the particle always visits all the three minima as they are all of the same energy.  (a'-d') $\eta<\eta_c$ case, three unfavored minima turn into maxima. The classical particle visit the minima in $\alpha=0$ but the trajectory gets farther from that point as the minumum becomes a maximum.}
    \label{fig_supplementary_traj}
\end{figure}
% {\Large P$\bar{3}1c$
% P$31c$
% R$\bar{3}c$
% R$3c$
% $\Gamma_2^-$
% $P_2$}

\end{document}